\documentstyle[11pt,aaspp4]{article}

\begin{document}
\title{The Multitude of Unresolved Continuum Sources  at 1.6 microns
in Hubble Space Telescope images of Seyfert Galaxies}

\author{
A.~C.\ Quillen\altaffilmark{1},
Colleen McDonald,
A.\ Alonso-Herrero, 
Ariane Lee,
Shanna Shaked, 
M.~J.\ Rieke, \& 
G.~H.\ Rieke,
}
\affil{Steward Observatory, The University of Arizona, Tucson, AZ 85721}
\altaffiltext{1}{aquillen@as.arizona.edu} 

\begin{abstract}
We examine 112 Seyfert galaxies observed by 
the Hubble Space Telescope (HST) at $1.6 \micron$.
We find that $\sim 50\%$ of the Seyfert 2.0 galaxies 
which are part of the Revised Shapeley-Ames (RSA) Catalog or 
the CfA redshift sample contain unresolved continuum 
sources at $1.6 \micron$.
All but a couple of the Seyfert 1.0--1.9 galaxies display unresolved
continuum sources.  The unresolved sources have fluxes
of order a mJy, near-infrared luminosities of order $10^{41}$ erg/s
and absolute magnitudes $M_H \sim -16$.
Comparison non-Seyfert galaxies from the RSA Catalog display significantly fewer
($\sim 20\%$), somewhat lower luminosity nuclear sources,
which could be due to compact star clusters.
We find that the luminosities of the unresolved Seyfert 1.0-1.9 sources
at $1.6\micron$ are correlated with [OIII]$5007\AA$
and hard X-ray luminosities, implying that these sources are non-stellar.
Assuming a spectral energy distribution similar to that
of a Seyfert 2 galaxy, we estimate that a few percent 
of local spiral galaxies 
contain black holes emitting as Seyferts at a moderate fraction, 
$\sim 10^{-1}$--$10^{-4}$, of their Eddington luminosities.

We find no strong correlation between $1.6\micron$ fluxes and
hard X-ray or [OIII] 5007$\AA$ fluxes for the pure Seyfert 2.0
galaxies. These galaxies also tend to have lower $1.6\micron$
luminosities compared to the Seyfert 1.0-1.9 galaxies of similar
[OIII] luminosity.
Either large extinctions ($A_V \sim 20-40$) are present towards their
continuum emitting regions or some fraction of the unresolved
sources at $1.6\micron$ are compact star clusters.
With increasing Seyfert type
the fraction of unresolved sources detected at $1.6\micron$ and
the ratio of $1.6\micron$ to [OIII] fluxes tend to decrease.
These trends are consistent with the 
unification model for Seyfert 1 and 2 galaxies. 
\end{abstract}

\section {Introduction}

Studies of active galactic nuclei (AGNs) 
have often focused on high luminosity objects
since in these objects the active nucleus
dominates the emission of the host galaxy.  
Study of the lower luminosity objects is 
often hampered by confusion with emission from the galaxy
in which the AGN resides (e.g., \cite{edelson}; \cite{spinoglio},
\cite{fadda}; \cite{almu96}). 
However, the high angular resolution of the Hubble Space Telescope (HST)
allows us to probe the nuclei with a beam area about 30
times smaller than is typically achieved with 
ground-based observations at these wavelengths.  
This enables us to separate the nuclear emission from that of
the surrounding galaxy with unprecedented accuracy.
\cite{malkan_3} have carried out a survey of nearby
Seyfert galaxies using WFPC2 on board HST at $0.6\micron$.
In this work, unresolved continuum sources 
(e.g., \cite{malkan})
were detected almost exclusively in Seyfert 1 galaxies.
These authors postulated that extinction associated
with a central torus (e.g., \cite{antonucci}) or on larger scales 
makes it difficult to detect nuclear sources associated with
Seyfert 2 galaxies.

Because extinction is comparatively reduced at longer wavelengths,  
the dusty torus model unifying Seyfert 1 and 2 galaxies
suggests that we should detect
nuclear emission from a larger fraction of Seyfert galaxies 
in the near-infrared than is possible at visible wavelengths.
Near-infrared ground based studies have detected 
bright non-stellar unresolved nuclear sources 
in a few bright Seyfert 2 galaxies
(e.g., \cite{malkan83}; \cite{almu96}), implying that
the extinction at this wavelength can be low enough
for continuum radiation to escape the central region.
Here we report on a survey of Seyfert galaxies observed with NICMOS
(the Near Infrared Camera and Multi-Object Spectrograph) on board
HST at 1.6$\micron$.   By using NICMOS, we combine the high
angular resolution of HST with the ability to carry out
an imaging survey in the near-infrared. 


\section{Archival Observations}

We compiled images from the HST archive that were
observed with the F160W filter at $1.6\micron$ with NICMOS.
These galaxies were observed 
primarily as part of three observing programs
which we identify by the proposal ID number
used by the Space Telescope Science Institute.
Galaxies from proposal 7330 were drawn from the Revised Shapely-Ames (RSA) 
Catalog ($B_T < 13.4$; \cite{RSA}) and are described by \cite{regan99_}.
This proposal includes a comparison sample of non-active
galaxies matching in luminosity, Hubble type, color and redshift 
distribution to its Seyfert sample.
Those from proposal 7328 are Seyfert galaxies with redshifts less 
than 0.019 from \cite{veron_}.
Those from proposal 7867 are  the 23 Seyfert 1.8-2 galaxies from
the CfA redshift survey (excluding NGC 1068 which was 
a GTO target) and are described by \cite{martini}.
In total we find 35 Seyfert galaxies identified in
the CfA redshift survey 
(e.g., \cite{huchra}; \cite{osterbrock}) 
including NGC 1068 and about 10 Seyfert 1.0-1.5 galaxies.
A total of 26 galaxies were listed as Seyfert galaxies 
in the survey by \cite{ho_3}, and
57 galaxies are part of the extended RSA sample discussed by 
\cite{maiolino95_}.
The galaxies are listed in Tables 1-5.

The CfA sample is drawn from the fraction of the sky
defined either by
$\delta \ge 0^\circ$   and $b \ge 40^\circ$ or
$\delta\ge -2^\circ.5$ and $b \le-30^\circ$.
Because it is not a color selected sample, it should
be relatively free of selection effects that tend to enhance
the proportion of galaxies with anomalously strong emission in the color 
used for selection (\cite{huchra}; \cite{osterbrock}).
However, because many of the objects are moderately distant,
the CfA sample does not sample the low luminosity 
tail of the Seyfert distribution (\cite{mcleod}).
It also does not contain enough Seyferts (only 51) to allow strong
statistical tests.  The RSA sample includes   
galaxies all over the sky.  The primary selection criterion is that
$B_T < 13.4$.  The mean distance of this sample is $D = 34$ Mpc,
about 3 times nearer than the CfA sample.  Nuclear spectra 
are less contaminated by galaxy light and Seyferts at a
larger range of galaxy inclinations and Hubble types are found in
this sample (\cite{maiolino95}). 
Unfortunately the spectroscopic identifications 
were not done with uniform data.  The more uniform spectroscopic survey
of \cite{ho_}, also based on the RSA catalog but not covering the
whole sky, has found a few additional low luminosity Seyfert galaxies
which were not compiled by \cite{maiolino95_}. 
\cite{ho_} also discovered some broad line components 
not previously seen with lower quality spectra.

We group the Seyfert galaxies according to samples
discussed in the literature.  Table 1 contains
all the galaxies which were part of the extended RSA 
sample (\cite{maiolino95}).  
Subscripts are given to Seyferts which
are also part of the CfA sample (e.g., \cite{osterbrock})
or which were observed by \cite{ho_}.
Galaxies which were not listed by \cite{maiolino95_} but
are contained in the CfA sample are included in Table 2.
Additional Seyferts are listed in Table 3.
Non-Seyfert galaxies are listed in Tables 4 and 5.

Images were reduced with the `nicred' data reduction
software (\cite{mcl}) using on orbit darks
and flats.  Each set of images 
was then combined according to the position observed.
The pixel size for the NICMOS camera 2 is $\sim 0''.076$
and for camera 1 is $\sim 0''.043$.
The FWHM for an unresolved point source is $\sim 0''.13$
at $1.6 \micron$ with HST corresponding to $\sim 20$ pc 
for the mean galaxy at a distance of 33 Mpc in the RSA sample.
Almost all of the images were observed with the sequence
of non-destructive reads in the MULTIACCUM mode.
We found no evidence for saturation in any of the images.

\section{Unresolved nuclear sources}

At the center of these galaxies we expect
contribution from both an underlying stellar component
and an unresolved nucleus.
To measure the flux from the unresolved component
we must subtract a resolved galaxian model.
We opted to use exponential and power law galaxian profiles since 
\cite{carollo98} find little or no morphological/photometric evidence for
a smooth, $R^{1/4}$ law bulge in WPFC2 images of galaxy bulges.
Since we fit the galaxy profile to the central arcsecond only,
a profile with more free parameters is not required.

For each camera we measured a point spread function from stars
in the images.   We constructed a library of galaxy profiles
for different scale lengths, $h$, or exponents, $\alpha$, by convolving 
the point spread function with exponential profiles 
(surface brightness $\propto e^{-r/h}$) 
or powerlaw profiles (surface brightness $\propto r^{-\alpha}$).
We then fit the sum of a convolved galaxian profile 
and the point spread function to the galaxy surface brightness
profiles.    
When the exponential profile was fit we 
varied the central surface brightness,
the scale length and the flux of an additional unresolved component.
When the powerlaw profile was fit we 
varied the surface brightness at a radius of $1''$,
the exponent, $\alpha$, and the flux of an additional unresolved component.
We also fit both powerlaw and exponential 
profiles without unresolved components to each galaxy.
We then identified the best fitting profile shape.
Sample fits to the galaxy
surface brightness profiles are shown in Figure 1.
For these galaxies we checked that the estimated unresolved flux 
was not strongly dependent upon the range of radius fit.
Doubling the fitting 
radius affected the estimated nuclear flux by less than 1\%.
In NGC 5252 we also fit the profile out to a radius of $5''$
with a Nuker law profile (a double powerlaw as described by \cite{faber}) 
and measured a nuclear flux that was only 10\% higher than that
found with a single powerlaw profile and a fit within $1''$.
Parameters describing the best fitting profiles
are listed in Tables 1-4.  
The error of the flux from the unresolved component was estimated
from the difference between the exponential and powerlaw profile fit.
When the best fitting profile contained no unresolved component
we then used the best fitting profile with an unresolved component
to derive an upper limit on the flux of a possible additional unresolved
source.

At $1.6 \micron$ an unresolved (point) source 
observed with NICMOS shows a prominent 
diffraction ring with a radius of $\sim 0''.3$.
This is the dominant feature we fit in the surface brightness profiles.
The error in this procedure we estimated from the scatter
in the residuals and was about $\pm 10\%$ of the measured flux for
the bright sources and about 50\% for most of the fainter sources
and is highest in the images with bright compact 
underlying surface brightness profiles.
To test our fitting procedure we recovered fluxes at these
levels of accuracy from model images created with
the iraf routine `mkobject'.   
For galaxies with extremely steep surface brightness profiles the
results of the fit are necessarily not unique.
For $h \gtrsim 0.''1$ and $\alpha \lesssim 1.3$ no diffraction
ring is seen clearly in the surface brightness profile of a model image
after convolution with the point spread function.
This describes the region in parameter space where the fitting 
procedure becomes uncertain.  In other words, for steep galaxy profiles
we cannot tell the difference between the sum of a point
source and a exponential at $h\sim 0''.1$
and an exponential profile with a similar scale length.

We list in tables 1-4 a symbol describing the morphology
of the central arcsecond.
When the unresolved point source dominated
the image we denoted in Tables 1-4 a type `*'.  When the
diffraction ring was faint but seen both
visually in the image and in the surface brightness profile
we denote `F'.  When no diffraction ring was seen but the
surface brightness profile was consistent with the sum of 
of an unresolved nuclear component and a smoother resolved exponential
profile we denote `:'.
When the nuclear profile was resolved we denote `-'.

The flux of the nuclear source was corrected using
aperture corrections derived from a point spread function
which we generated with Tinytim (\cite{krist}).
To convert fluxes into Jy we used  conversion factors
$2.360\times10^{-6}$, $2.190\times10^{-6}$, $2.776\times10^{-6}$
Jy per DN/s for Cameras 1, 2 and 3 respectively.
This flux calibration is based on
measurements of the standard stars P330-E and P172-D during
the Servicing Mission Observatory Verification program
and subsequent observations (M.~Rieke 1999, private communication).


\subsection{The fraction of galaxies with unresolved emission}

In Table 6 we compile the fraction of various Seyfert type galaxies
which display unresolved nuclear sources.
All but one (NGC 4594) of the Seyfert 1.0-1.9 galaxies (types
listed by \cite{maiolino95} and \cite{osterbrock}) displayed an unresolved 
nuclear source.   Additionally two galaxies
listed as S1.9's by \cite{ho} (NGC 2639 and NGC 4258) did not display unresolved emission.
About 50\% of the Seyfert 2.0 galaxies displayed
unresolved sources.  The CfA and Ho samples have the lowest fraction
of unresolved nuclear sources among the Seyfert 2.0 galaxies possibly
because they are the most uniform
in spectral quality and Seyfert typing 
compared to the other samples.
The presence of weak broad line emission may
indicate that continuum radiation at $1.6\micron$ can escape
the central pc.

About 50\% of the Seyfert 2 galaxies from proposal 7330 
displayed significant unresolved emission compared to 
24\% of the control or non-Seyfert galaxies drawn from this same proposal.
The fraction of galaxies with more robustly identified
unresolved sources (those labelled `*' and `F') is also larger
in the Seyfert sample than the non-Seyfert sample.
Though our fitting routine is not unique, particularly
when the galaxy surface brightness profiles are steep, if we
exclude the more marginal cases, we still find that 
the Seyfert and non-Seyfert samples differ.
This implies that the Seyfert galaxies are 
more likely to display unresolved nuclear sources at $1.6 \micron$
than the non-Seyfert galaxies.   
There must be an intrinsic difference
between galaxies identified as Seyferts and those not 
identified as Seyferts in the RSA Catalog.

HST observations of nearby spiral galaxies have found that many harbor
nuclear star clusters, a small fraction of which ($\sim 5\%$) are unresolved
(\cite{carollo97}).  
The non-Seyfert galaxies studied by \cite{carollo97_} 
are somewhat closer than the galaxies in our sample, lying at a
mean distance of 23 Mpc, compared to 34 Mpc, the average distance of 
our RSA sample. Using a color $V-H \sim 2.65$, the star 
clusters from \cite{carollo97_} have fluxes ranging from 0.01 to 5 mJy,
making them similar in magnitude to the unresolved fluxes we have measured in
the Seyfert sample.  Placing these galaxies at distances
similar to our Seyfert galaxies would result in a 
somewhat larger number of the sources
being unresolved, however, it would also decrease their brightness.

We can also compare the luminosity distributions of
the unresolved nuclear sources between the non-Seyfert and Seyfert 
RSA samples (galaxies observed as part of proposal 7330).  
We estimate the luminosity at $1.6\micron$ by
$\nu f_\nu  * 4 \pi D^2$ for a distance of $D$.
We used distances from the Nearby Galaxies
Catalog (\cite{tully}) or from the radial velocity with a Hubble constant
of 75 km~s$^{-1}$ Mpc$^{-1}$.

We find that the luminosity distribution
of unresolved sources in proposal 7330 and the RSA sample 
differ, even though their redshift
distributions are similar (see Figure 2).  
The non-Seyfert unresolved sources tend
to have lower luminosities.
We conclude that the luminosities and fraction of 
unresolved nuclear sources in Seyferts galaxies differ from 
those found in non-Seyfert galaxies.

\subsection{Comparison with hard X-ray and [OIII]5007$\AA$}

Mid-infrared photometric and optical and UV
spectroscopic surveys have found that Seyfert 2 galaxies are more likely to 
harbor nuclear star formation 
(\cite{gonzalez}; \cite{ruiz});
This might suggest that a galaxy identified as a Seyfert
2 galaxy may be more likely to harbor a brighter compact star cluster
than a non-Seyfert galaxy.  In this case 
the unresolved sources in the Seyfert 2 galaxies 
could be due to compact star clusters
rather than non-stellar AGN emission. 

There are a couple of ways to test this hypothesis.  
One way is to search for 
variability in multi-epoch observations.
\cite{shaked_} found that 
8/13 of the unresolved sources in Seyfert 1.8 and 1.9 galaxies 
varied, proving that they 
are non-stellar, associated with the central pc of an AGN, and not
emission from bright nuclear stellar clusters. 
However the $1.6\micron$ unresolved emission in  
Seyfert 2.0 galaxies lacking any broad line component
could still arise from compact star clusters.  Another way
to test this hypothesis is by searching for correlations between
the $1.6\micron$ emission, and 
[OIII] or hard X-ray emission.  Previous studies
have shown that [OIII] and hard X-ray luminosities are correlated
with AGN activity  (\cite{mulchaey94}; \cite{keel94}, \cite{bassani}),
and that the near infrared flux is correlated with hard X-ray and
[OIII] luminosity in bright Seyferts (\cite{almu96}; \cite{almu97}).
In Figure 3 we compare luminosities computed at $1.6 \micron$ 
with those estimated with [OIII]$5007\AA$ and hard X-ray fluxes.
[OIII]$5007\AA$ fluxes were taken from Ho et al.~(1995), Whittle (1992),
Bassani et al.~(1999) and Risaliti et al.~(1999) 
and whenever possible are corrected
for redenning using the Balmer decrement.
Hard X-ray fluxes (2-10keV) were taken from the
compilations of Bassani et al.~(1999), Risaliti et al.~(1999),
and Mulchaey et al.~(1994) and were corrected for observed absorption
when the sources were not Compton thick.  As discussed in these compilations 
the fluxes have been measured with a variety of different instruments, 
apertures and calibration techniques.
Some of the scatter in these plots may be due to inconsistencies between
the comparative measurements of the [OIII] or hard X-ray fluxes.

We computed Spearman rank-order correlation coefficients 
on the fluxes for the Seyferts shown in Figure 3 (see Table 7).
There is a convincing correlation between [OIII] and $1.6\micron$
fluxes and between hard X-ray and $1.6\micron$ fluxes in the Seyfert
1.0-1.9 galaxies.  However, only a weak correlation is seen 
in the pure Seyfert 2.0 galaxies.  This lack of a strong correlation
suggests that extremely large extinctions are present towards
the Seyfert 2.0 nuclei.   Alternatively some fraction
of the unresolved $1.6\micron$ Seyfert 2.0 sources 
could be due to star clusters.

Starburst galaxies are emitters of hard X-rays 
which could be arising from hidden AGN, high mass X-ray binaries,
or inverse Compton scattering from high energy particles
associated with supernovae (e.g., \cite{ohashi}).
For comparison to our galaxy nuclei,
we estimate the $1.6\micron$ to hard X-ray
luminosity ratio for the infrared luminous starburst galaxy NGC 3256.   
NGC 3256 has a hard X-ray (2-10 keV)
luminosity of $2 \times 10^{41}$ erg/s (\cite{moran}) and
a luminosity at $1.6\micron$ of $3.4\times 10^{43}$ erg/s
which we estimate
based upon the H band aperture photometry of \cite{glass}.
The ratio of the $1.6 \micron$ to hard X-ray luminosity
is 170 (a log of 2.2) which is above the Seyfert 2 points shown in Fig.~3b.  
The bulk of the hard X-ray emission in the Seyfert 2s must
come from an AGN rather than a nuclear star cluster.  
If the hard X-ray emission came from a starburst then
we would have expected even larger levels of soft
X-ray emission which is generally not seen in Seyfert 2 galaxies
(e.g., \cite{mulchaey94}).

We see in Figure 3 that for a given [OIII] luminosity the Seyfert 1.8-1.9
galaxies have $1.6 \micron$ luminosities similar to or slightly lower 
than the Seyfert 1.0--1.5 galaxies.
The Seyfert 2.0 galaxies, however, have lower $1.6\micron$ fluxes.
If the 1.6 $\micron$ emission is associated with AGNs then 
there could be significant extinction, $A_V \sim 40$, 
towards the continuum emission region in the Seyfert 2 galaxies.
A similar trend was observed in a smaller ground based sample 
by \cite{almu97_3}.
If the [OIII] luminosity is a reliable luminosity indicator, then 
Seyfert 2 galaxies have significantly larger extinctions towards
their continuum emission regions than Seyfert 1.0--1.9 galaxies.
The Seyfert 1.8--1.9 galaxies appear to be intermediate,
suggesting that a partial view of the broad line region 
occurs when there is reduced extinction towards the 
near-infrared continuum emission region.

The star formation models of \cite{fioc} predict that 
a $10^6$ year old solar metallicity instantaneous starburst 
should have a ratio of [OIII] to $1.6\micron$
luminosity of 2.7.  For older starbursts the [OIII]
luminosity drops rapidly compared to the 1.6 micron luminosity
(at $10^7$ years the ratio is $\sim 0.002$).
The emission line ratios of these galaxy nuclei have caused them to 
be identified as Seyferts, so it is unlikely that most of the [OIII]
emission arises from a starburst (though starburst models
for Seyfert line ratios have been proposed; \cite{terlevich}).  
However, we can consider what
minimum starcluster ages could be consistent with a [OIII] contribution
from the starburst contribution that does not dominate the emission line 
spectrum.

The [OIII] luminosities of the Seyfert 1.0-1.9 galaxies are well
above that predicted from a young starburst.
However the [OIII] to $1.6 \micron$ luminosity ratios of the Seyfert 2
galaxies are similar to those observed in $10^6$ year old starbursts.  
It is unlikely that the $1.6 \micron$ emission in the Sy 2 galaxies 
is associated with a $10^6$ year old starcluster, because then
the nuclear spectrum would be that of an HII region rather
than a Seyfert.
However older few million year old starclusters cannot be excluded.
In this case the $1.6\micron$ emission could be stellar and the
narrow emissions lines associated with the AGN.

\subsection{Minimum foreground extinctions}

The spectral energy distribution of an old stellar population
peaks at about $1.6 \micron$.  
If the spectral energy distribution of a continuum source
associated with an AGN is similar to that of a quasar, which
generally has a dip at $1.6 \micron$, then it should be
the {\it most } difficult to detect 
against the background galaxy at this wavelength.
Likewise Seyfert 1 galaxies have continua 
which are bluer than an old stellar population so
we might expect that Seyfert 1 galaxies would be more 
difficult to detect at $1.6 \micron$ than in the visible bands.
The ``unification'' model postulates that 
Seyfert 1 and 2 galaxies differ
in terms of orientation angle (\cite{antonucci}), and that
a dusty torus absorbs a significant fraction of the optical/UV/X-ray
luminosity.  This implies that significant
extinction in front of the nucleus may be present 
in Seyfert 2 galaxies.  This extinction may account for the large number
of unresolved point sources detected at $1.6 \micron$ compared
to the non-detections reported by \cite{malkan_} at $0.606 \micron$.

The galaxies chosen by \cite{malkan_} for WFPC2 observations 
were galaxies from the Catalog of Quasars
and Active Nuclei (\cite{veron}) with redshift less than  $0.035$.
Galaxies observed with NICMOS as part of proposal 7328 were also
chosen from this catalog, (restricted to $z< 0.019$), however most of the 
galaxies observed with NICMOS were not.  
Of the galaxies which were part of the 7328 proposal we detected
unresolved emission from all Sy 1-1.9s (12 Sy 1.0-1.5s and 4 Sy 1.8,1.9s).
Out of 13 Sy 2 galaxies, unresolved emission was detected from 8 at high
confidence and from 1 with lower confidence.  This fraction
of Sy 1.8-2.0 with unresolved nuclear emission is larger than
that reported by \cite{malkan_}.
Unfortunately, some of the WFPC2 images of this survey
were saturated near the galaxy nuclei.  Images observed
with shorter exposure times might have displayed a larger number of unresolved
nuclear sources.

We can assume that the level of stellar emission limits our ability
to detect the non-stellar emission.
An old stellar population commonly found in the central region of a
galaxy has $V-H  \sim 3.0$ (\cite{frogel}).
If Seyfert 2 galaxies are similar to Seyfert 1 galaxies
in their inner regions
then we can model the underlying emission as that of  a
Seyfert 1 with $V-H \sim 2.6$ (\cite{alloin}).
For the non-stellar source to be detectable at $1.6\micron$
(roughly H band) and undetectable at $0.6\micron$
(roughly V band), the source must be redder than 
the stellar background.    If we assume that foreground extinction
is responsible for the redenning 
then the change in color  must be larger than the difference between
the nucleus and stellar color ($\Delta (V-H) \gtrsim 0.4$).
Using $A_H \sim 0.176 A_V$
(from \cite{mathis})
$\Delta (V-H) \sim A_V - A_H \sim 0.824 A_V$
which implies that foreground extinction at least 
$A_V \gtrsim 0.5$ is needed for the unresolved sources
to be detectable at $1.6\micron$ and not at $0.6\micron$
against the stellar background.
Some of the unresolved sources listed in Tables 1-3 are 
up to 100 times
above the detection limit at $1.6\micron$ (e.g., NGC~1068)
which implies that
extinctions of at least $A_V \gtrsim 5$ are required to account for 
their brightness at $1.6\micron$ and faintness at visible wavelengths
which would be consistent with a reduced detection rate in
the optical survey (\cite{malkan}).

\section{Summary and Discussion}

We report on the discovery of a large number of unresolved continuum 
emission sources at $1.6\micron$ in a significant fraction 
of nearby Seyfert galaxies observed with HST.
Of the Seyfert 2 galaxies in the RSA and CfA samples
50--70\% display unresolved continuum sources.
For Seyfert 2.0 galaxies listed in \cite{ho_} only 10--35\% of the Seyfert
2.0 galaxies displayed unresolved sources.  
All but 1 of the Seyfert 1.0-1.9 galaxies display 
unresolved sources.
A comparison galaxy sample drawn from the RSA Catalog lacking
Seyfert nuclei display significantly fewer ($\sim 20\%$) unresolved
sources than Seyferts found in this catalog.  
We find that the luminosities and fraction of
unresolved nuclear sources in Seyfert galaxies differ
from those found in non-Seyfert galaxies.

The luminosities at $1.6\micron$ are correlated with 
hard X-ray and [OIII] 5007$\AA$ luminosities for the Seyfert 1.0-1.9
galaxies.  These unresolved sources are therefore most likely 
non-stellar and not due to compact nuclear star clusters.
The presence of weak broad
line emission (in Seyfert 1.8 and 1.9 galaxies) 
appears to be coincident with the presence of a detectable 
unresolved continuum source at $1.6\micron$.
This is surprising since the size of the broad line region
is expected to be much smaller than that containing the
hot dust giving rise to the near-infrared emission (e.g., \cite{barvainis};
\cite{pier}; \cite{marconi}).
A partial covering of the broad line region may be directly 
associated with reduced extinction towards the 
near-infrared continuum emitting region.
The near-infrared continuum emission region could be closer
to the broad line region than previously considered.

We find no strong correlation between $1.6\micron$ fluxes and
hard X-ray or [OIII] 5007$\AA$ fluxes for the pure Seyfert 2.0.
These galaxies also tend to have lower $1.6\micron$ 
luminosities compared to the Seyfert 1.0-1.9 galaxies of similar
[OIII] luminosity. 
Either large extinctions ($A_V \sim 20-40$) are present towards their
continuum emitting regions or/and some fraction of the unresolved
sources at $1.6\micron$ are compact star clusters.
With increasing Seyfert type 
the fraction of unresolved sources detected at $1.6\micron$ and
the ratio of $1.6\micron$ to [OIII] fluxes tend to decrease.
These trends are consistent with
the unification model for Seyfert 1 and 2 galaxies.

Assuming a color typical of a 
Seyfert 1 galaxy, only a moderate amount
of foreground extinction, $A_V \gtrsim 0.5$, is required to account
for the detections at $1.6\micron$ and non-detections at $0.6\micron$
(reported by \cite{malkan}) of the Seyfert 1.8-2.0 galaxies.
We suspect that an even larger number of galaxies would
display unresolved sources at longer wavelengths if observed
at a similar angular resolution.

Accretion models for AGNs rely on two fundamental parameters
to describe them, the black hole mass and the bolometric
luminosity emitted,  which we expect is dependent upon the accretion
rate.    Black hole masses have recently been measured
with a variety of techniques (e.g., \cite{richstone}), however
estimates of the bolometric luminosity exist for only a few
nearby sources.
We can crudely estimate the bolometric luminosity 
from that at $1.6\micron$ by assuming a ratio of $\sim 10$ between
the $1.6\micron$ and mid-IR luminosity similar to that
of Seyfert 2 galaxies (e.g, \cite{fadda} and
a ratio of $\sim 10$ between the mid-IR and bolometric luminosity 
(e.g., \cite{spinoglio}).
In units of the Eddington luminosity
%
\begin{equation}
{L_{bol} \over L_{ED}} \sim 10^{-2} 
  \left({L_{1.6\micron} \over 10^{41}{\rm ergs/s}}\right)
  \left({L_{bol}/ L_{1.6 \micron} \over 100}\right)
  \left({M_{BH} \over 10^{7}M_\odot}\right)^{-1}
\end{equation}
About $10\%$ of the RSA sample of galaxies contains Seyfert nuclei
with Seyfert 1.8-2 galaxies being 3 times more common than Seyfert 1-1.5
galaxies (\cite{maiolino95}).  Our RSA subsample
contains a substantial fraction of Seyfert galaxies
with $1.6\micron$ luminosities of order $10^{41}$ ergs/s
(see Figure 2).
Since most of the galaxies are spiral galaxies 
we expect black hole masses in the range of $10^6 - 10^8 M_\odot$
(e.g., \cite{richstone}).
The above estimate implies that the bolometric luminosities in Eddington units 
span the range $10^{-1}-10^{-4}$
for black holes likely to reside in these galaxies.
This range is consistent with previous estimates 
(e.g., \cite{wandel}; \cite{cavaliere}),
and suggests that a few percent of the black holes 
resident in local spiral galaxies are emitting as Seyferts at a moderate
fraction of their Eddington luminosity.
Longer wavelength observations will yield better estimates
for the bolometric luminosities of these numerous 
low luminosity AGNs.

The Seyfert samples we have considered in this paper consist
of Seyfert galaxies with clear optical spectroscopic identifications.
A sample of active galaxies chosen in the mid-infrared (e.g.~10-50 microns)
or with hard X-rays may yield a population of more highly obscured
AGNs which may be much harder to detect both optically
and at $1.6\micron$.

\acknowledgments

%
Support for this work was provided by NASA through grant number
GO-07869.01-96A
from the Space Telescope Institute, which is operated by the Association
of Universities for Research in Astronomy, Incorporated, under NASA
contract NAS5-26555.
We also acknowledge support from NASA project NAG-53359 and
NAG-53042 and from JPL Contract No.~961633.
This research has made use of the NASA/IPAC Extragalactic Database (NED) 
which is operated by the Jet Propulsion
Laboratory, California Institute of Technology, under contract with 
the National Aeronautics and Space Administration. 
We thank M.~Salvati, and K.~Gordon for helpful discussions.

\clearpage


\vfill\eject

\begin{figure*}
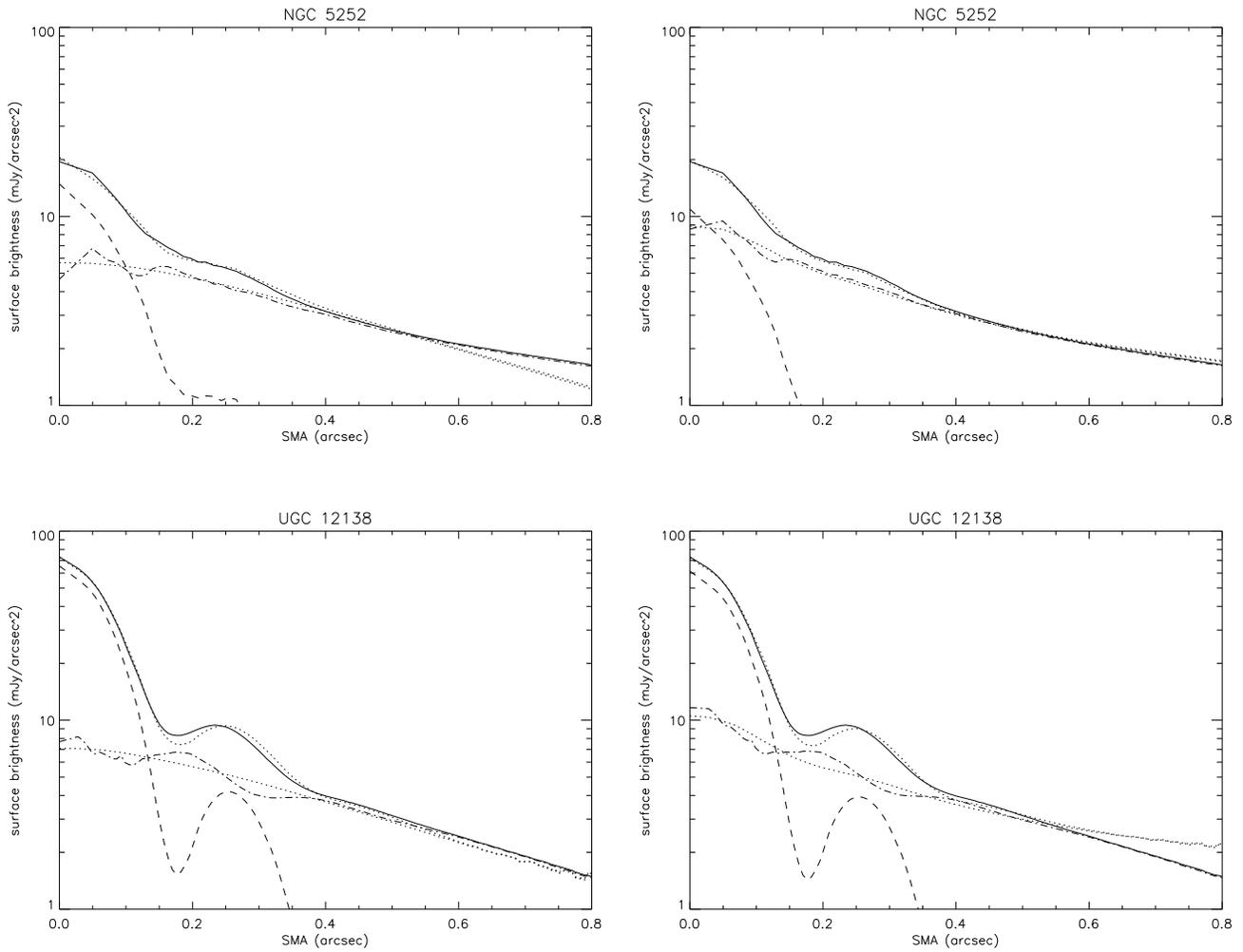

\caption[junk]{
Examples of fits done to the galaxy surface brightness profiles
to allow measurement of unresolved sources.
The upper solid line is the galaxy profile and the upper dotted
line is the resulting fit to this profile.  The fit is a sum
of a point source (shown as the dashed line) and an exponential
profile or powerlaw profile which has been convolved with the point 
spread function 
(shown as the lower dotted line).  The lower dot dashed line
is the galaxy profile subtracted by the point source.
The point spread functions shown were measured from stars observed
in the same filter and with similar exposure times.
a) NGC 5252 was observed with NICMOS Camera 2. Its profile is 
fit with an exponential galaxian profile.
b) NGC 5252 profile fit with a powerlaw galaxian profile.
c) UGC 12138 was observed with NICMOS Camera 1 and fit
with an exponential galaxian profile.
d) UGC 12138 fit with a powerlaw galaxian profile.
}
\end{figure*}

\begin{figure*}
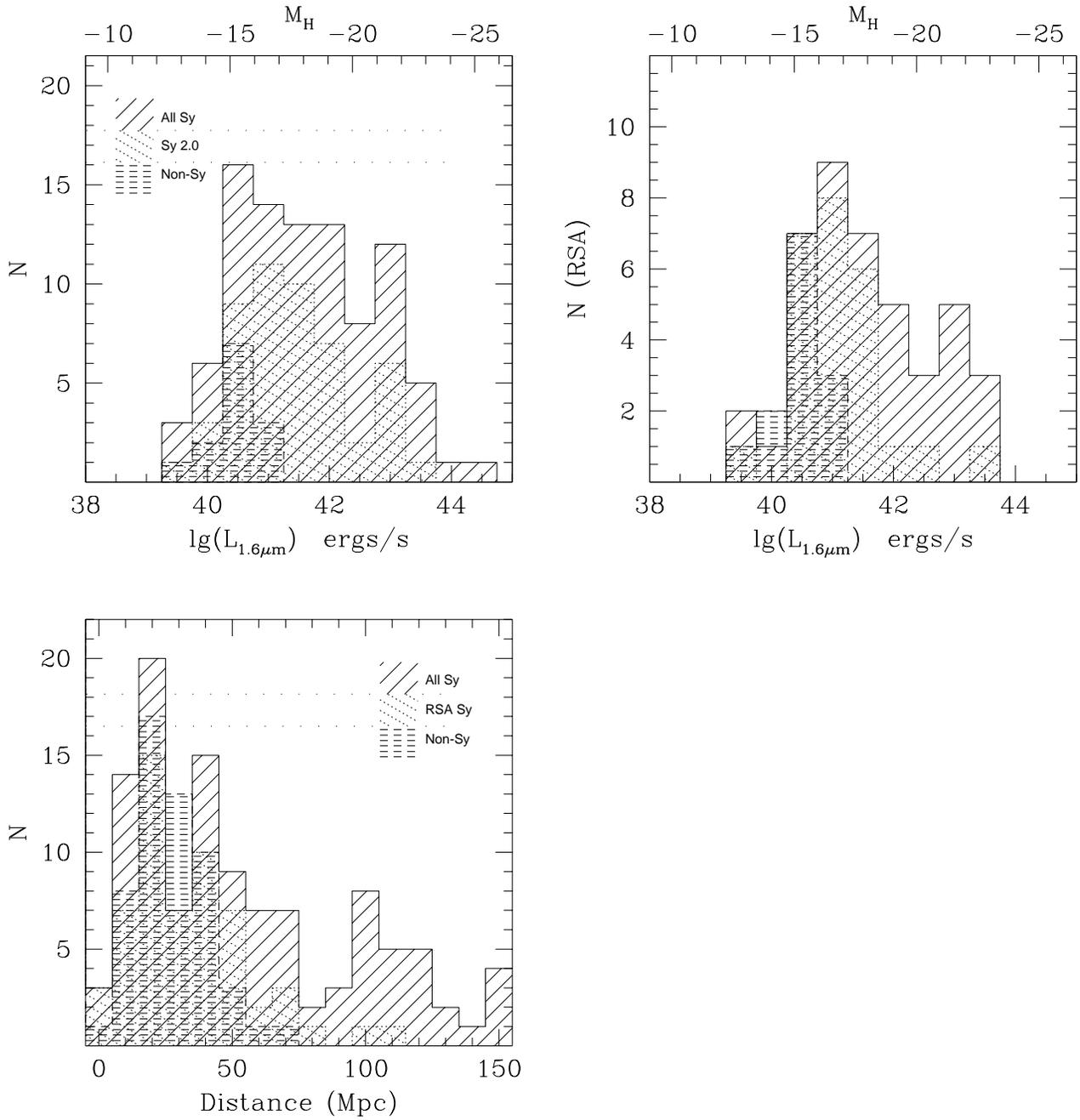

\caption[junk]{
a) Luminosity distribution for the Seyferts with unresolved sources.
The histogram filled with solid lines shows all the Seyferts 
listed in Tables 1-3,
and the histogram filled with dotted lines corresponds to the 
Seyfert 2.0 galaxies. 
The histogram filled with dashed horizontal lines 
corresponds to the unresolved
sources in the non-Seyfert, control galaxies.  Absolute H magnitudes are shown
on the top of the histograms.
b) Luminosity distribution for the Seyferts in the RSA sample.
c) Distance distribution for the total Seyfert, the RSA 
Seyfert and non-Seyfert samples.
The distance distribution of the non-Seyfert galaxies matches that 
of the RSA sample.
}
\end{figure*}

\begin{figure*}
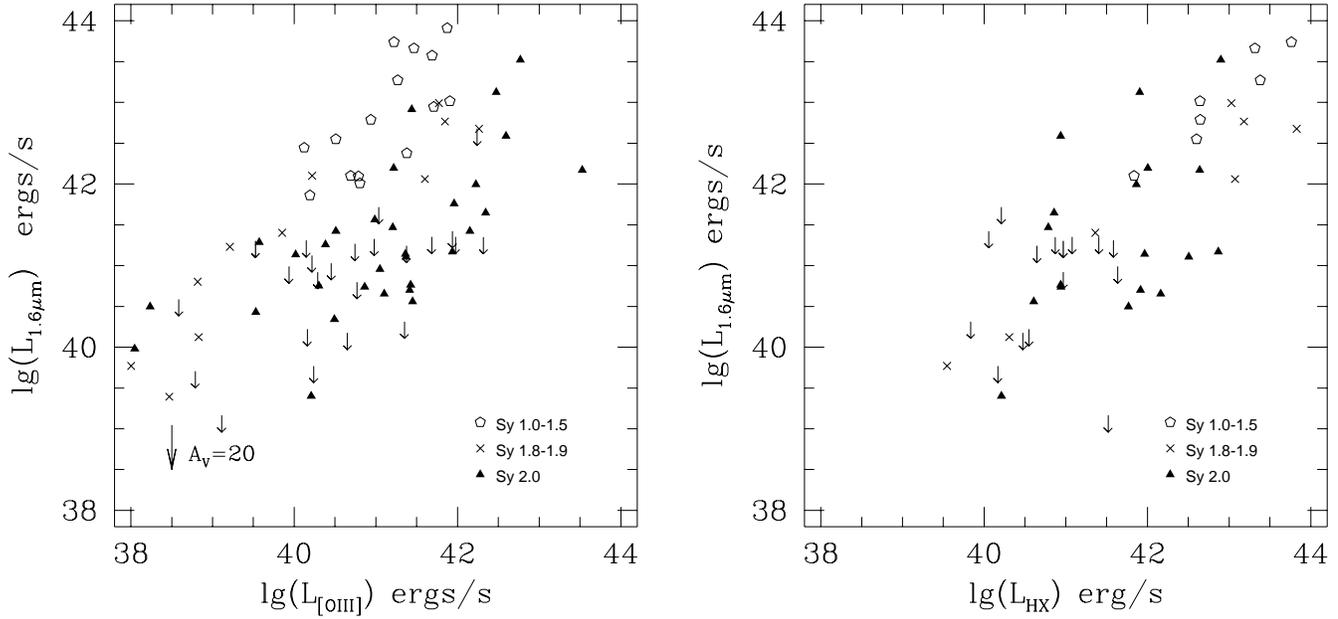

\caption[junk]{
a) Correlation between $1.6\micron$ and [OIII]$5007\AA$
luminosities for the Seyferts listed in Tables 1-3.
Upper limits are given when no nuclear point source was detected
at $1.6 \micron$.
Seyfert 2.0 galaxies appear to have weaker $1.6 \micron$ luminosities
compared to Seyfert 1.9-1.0 galaxies. 
The correlation between [OIII] and $1.6\micron$ luminosity
suggests that the majority of the unresolved sources are non-stellar.
[OIII]$5007\AA$ fluxes were taken from Ho et al.~(1995), Whittle (1992),
Bassani et al.~(1999) and Risaliti et al.~(1999)
and whenever possible are corrected
for redenning using the Balmer decrement.
Points or upper limits are shown only when we found 
[OIII] fluxes in these compilations.
b) Correlation between $1.6\micron$ and hard X-ray luminosities.
Hard X-ray fluxes (2-10keV) were taken from Bassani et al.~(1999),
Risaliti et al.~(2000)
and Mulchaey et al.~(1994) and were corrected for observed absorption,
though some of the Seyfert 2.0 galaxies are Compton thick and so
the fluxes do not represent the true X-ray luminosities.
Points or upper limits are shown only when we found
hard X-ray fluxes in these studies.
}
\end{figure*}

\end{document}


\pagestyle{empty}

\begin{deluxetable}{lccccccccccc}
\tablewidth{0pt}
\tablecaption{RSA Seyfert Sample}
\tablehead{
\multicolumn{1}{l}{Galaxy}     &
\colhead{Nucleus}    &
\colhead{Type}       &
\colhead{Proposid}   &
\colhead{$v_{hel}$}  &
\colhead{profile}    &
\colhead{$f_{nuc}$}  &
\colhead{$S_0|S_{r=1''}$}        &
\colhead{$h|\alpha$}  \\
\colhead{}                &
\colhead{}                &
\colhead{}                &
\colhead{}                &
\colhead{km/s}            &
\colhead{}                &
\colhead{mJy}             &
\colhead{mJy $''^{-2}$}     &
\colhead{}          \\
\colhead{(1)}             & 
\colhead{(2)}             & 
\colhead{(3)}             & 
\colhead{(4)}             & 
\colhead{(5)}             & 
\colhead{(6)}             & 
\colhead{(7)}             & 
\colhead{(8)}             & 
\colhead{(9)}              
} 
\startdata
NGC 788    &- & 2       & 7330/2 & 4078  & P &   $<0.2$            &   1.34 &   0.97 \\
NGC 1068$^{c,h}$&*&2/1.8& 7215/2 & 1137  & - &  83.6  $\pm$   8    &   -    &   -    \\
NGC 1241   &: & 2       & 7330/2 & 4052  & P &   0.17 $\pm$   0.07 &   2.49 &   0.41 \\
NGC 1275$^h$&*& 1.9/1.5 & 7330/2 & 5264  & - &   4.30 $\pm$   0.43 &   -    &   -    \\
NGC 1320   &* & 2       & 7330/2 & 2716  & P &   1.15 $\pm$   0.35 &   2.67 &   0.74 \\
NGC 1386   &- & 2       & 7458/2 & 868   & P &   $<0.2$            &   7.84 &   0.41 \\
NGC 1667$^h$&-& 2       & 7330/2 & 4547  & P &   $<0.3$            &   1.56 &   0.79 \\
NGC 2273$^h$&:& 2       & 7172/2 & 1849  & P &   0.32 $\pm$   0.28 &   3.50 &   0.61 \\
NGC 2639$^h$&-& 2/1.9   & 7330/2 & 3187  & E &   $<0.15$           &  21.82 &   0.32 \\
NGC 3031$^h$&*& 1.8/1.5 & 7331/2 & -34   & - &  13.4  $\pm$   1.3  &   -    &   -    \\
NGC 3081   &: & 2       & 7330/2 & 2385  & P &   0.22 $\pm$   0.13 &   1.96 &   0.67 \\
NGC 3227$^{c,h}$&*&1.5  & 7172/2 & 1157  & - &   13.2 $\pm$   1.3  &   -    &   -    \\
NGC 3362$^c$&F& 2       & 7867/1 & 8318  & P &   0.05 $\pm$   0.02 &   0.32 &   0.84 \\
NGC 3393   &- & 2       & 7330/2 & 3750  & P &   $<0.25$           &   2.38 &   0.73 \\
NGC 3516$^{c,h}$&*&1.2  & 7330/2 & 2649  & - &  18.1  $\pm$   1.8  &   -    &   -    \\
NGC 3786$^c$&*& 1.8     & 7867/1 & 2678  & E &   3.25 $\pm$   0.33 &  18.97 &   0.33 \\
NGC 3982$^{c,h}$&F&2/1.9& 7330/2 & 1109  & P &   0.34 $\pm$   0.11 &   1.29 &   0.65 \\
NGC 4151$^{c,h}$&*&1.5  & 7215/2 & 995   & - & 112.1  $\pm$  11    &   -    &   -    \\
NGC 4235$^{c,h}$&*&1.2  & 7328/2 & 2410  & P &   3.69 $\pm$   0.38 &   3.10 &   0.39 \\
NGC 4253$^c$&*& 1.5     & 7330/2 & 3876  & - &  20.0  $\pm$   2.0  &   -    &   -    \\
NGC 4258$^h$&-& 2/1.9   & 7230/2 & 448   & P &  $<1.0$             &   5.35 &   0.98 \\
NGC 4388$^{c,h}$&*&2/1.9& 7867/1 & 2524  & P &   0.71 $\pm$   0.22 &   1.61 &   0.80 \\
NGC 4395$^{c,h}$&*&1.8  & 7330/2 & 319   & - &   0.85 $\pm$   0.18 &   -    &   -    \\
NGC 4593   &* & 1       & 7330/2 & 2698  & - &  10.1  $\pm$   1.0  &   -    &   -    \\
NGC 4594$^h$&-& 1.9/L2  & 7331/2 & 1091  & P &    $<1.4$           &  30.28 &   0.38 \\
NGC 4785   &- & 2       & 7330/2 & 3735  & P &    $<0.3$           &   2.79 &   0.70 \\
NGC 4939   &F & 2       & 7330/2 & 3111  & P &   0.36 $\pm$   0.06 &   1.89 &   0.47 \\
NGC 4941   &- & 2       & 7330/2 & 1108  & P &    $<0.4$           &   3.10 &   0.71 \\
NGC 4945   &- & 2       & 7865/2 & 560   & E &    $<0.15$          &  21.27 &   0.41 \\ 
NGC 5005$^h$&-& 2/L1.9  & 7330/2 & 946   & P &    $<3.2 $          &   8.88 &   0.82 \\  
NGC 5033$^{c,h}$&*&1.9/1.5&7330/2& 875   & P &   3.22 $\pm$   0.46 &   4.97 &   0.64 \\
NGC 5128   &* & 2       & 7330/2 & 547   & - &   5.8  $\pm$   0.6  &   -    &   -    \\
NGC 5135   &* & 2       & 7330/2 & 4112  & P &   0.66 $\pm$   0.07 &   1.57 &   0.50 \\
NGC 5194$^h$&:& 2       & 7327/2 & 463   & E &   0.19 $\pm$   0.18 &  17.16 &   0.74 \\
NGC 5273$^{c,h}$&*&1.9/1.5&7330/2& 1089  & P &   1.67 $\pm$   0.17 &   2.18 &   0.45 \\
NGC 5347$^c$&*& 2       & 7330/2 & 2335  & P &   0.97 $\pm$   0.24 &   1.20 &   0.85 \\
NGC 5427   &* & 2       & 7330/2 & 2618  & P &   0.42 $\pm$   0.09 &   0.59 &   0.78 \\
NGC 5506   &* & 1.9     & 7330/2 & 1853  & - &  53.1  $\pm$   5.3  &   -    &   -    \\
NGC 5548$^{c,h}$&*&1.2/1.5&7172/2& 5149  & P &  17.7  $\pm$   1.8  &   1.18 &   0.50 \\
NGC 5643   &- & 2       & 7330/2 & 1199  & P &  $ <1.7$            &   2.60 &   1.09 \\
NGC 5674$^c$&*& 1.9  & 7867/1 & 7474  & E &   2.62 $\pm$   0.26 &  12.01 &   0.30 \\
NGC 5953   &F & 2    & 7330/2 & 1965  & P &   0.37 $\pm$   0.26 &   4.36 &   0.56 \\
NGC 6221   &* & 2    & 7330/2 & 1482  & P &   3.14 $\pm$   0.30 &   3.48 &   0.32 \\
NGC 6300   &* & 2    & 7330/2 & 1110  & P &   1.97 $\pm$   0.20 &   3.01 &   0.37 \\
NGC 6814   &* & 1.5  & 7330/2 & 1563  & P &   6.25 $\pm$   0.63 &   3.04 &   0.44 \\
NGC 6890   &* & 2    & 7330/2 & 2419  & E &   0.80 $\pm$   0.09 &   5.14 &   0.76 \\
NGC 7130   &- & 2    & 7330/2 & 4842  & E &   $<0.15$           &  17.22 &   0.32 \\
NGC 7469$^c$&*& 1.2  & 7219/2 & 4892  & - &  48.3  $\pm$   4.8  &   -    &   -    \\
NGC 7479$^h$&:& 2/1.9& 7331/2 & 2381  & P &   0.24 $\pm$   0.14 &   1.63 &   0.71 \\
NGC 7496   &- & 2    & 7330/2 & 1649  & P &   $<1.4 $           &   0.63 &   1.36 \\
NGC 7582   &* & 2    & 7330/2 & 1575  & P &  22.6  $\pm$   2.3  &   5.08 &   0.31 \\
NGC 7743$^h$&-& 2    & 7330/2 & 1710  & P &   $<0.5$            &   2.57 &   1.13 \\
IC 2560    &: & 2    & 7330/2 & 2925  & P &   0.16 $\pm$   0.09 &   2.99 &   0.42 \\
IC 5063    &* & 2    & 7330/2 & 3402  & P &   0.32 $\pm$   0.12 &   2.42 &   0.63 \\ 
CIRCINUS   &* & 2    & 7273/2 & 436   & P &   4.77 $\pm$   0.7  &  12.74 &   0.51 \\
MRK 1066   &F & 2    & 7330/2 & 3605  & P &   0.51 $\pm$   0.16 &   3.53 &   0.60 \\
IRAS1832-5926&*& 2   & 7328/1 & 6065  & - &  22.7  $\pm$   4.0  &   -    &   -    \\
        &  &      &        &       &         &          &       \\ 
        &  &      &        &       &         &          &       \\ 
        &  &      &        &       &         &          &       \\ 
        &  &      &        &       &         &          &       \\ 
        &  &      &        &       &         &          &       \\ 
        &  &      &        &       &         &          &       \\ 
\enddata
\tablenotetext{c}{This galaxy is also part of the CfA Seyfert
sample (Huchra \& Burg 1992; Osterbrock \& Martel 1993)}
\tablenotetext{h}{This galaxy is also part of the sample studied
by Ho, Filippenko \& Sargent (1995).  
}
\tablecomments{
This table contains galaxies which were part of the RSA Seyfert
sample listed by Maiolino \& Rieke (1995).  
Columns (1) Galaxy; 
(2) Type of nucleus seen in the F160W (1.6 micron) NICMOS images.  
When the nucleus
displayed a clear diffraction ring we denote '*', when the ring
was extremely faint we denote 
'F', and when the galaxy was resolved we denote '-'.  When there
was an unresolved peak but no sign of a diffraction ring we denote ':'.
(3) Seyfert type. 
These classifications are following those compiled from the literature
by Maiolino \& Rieke (1995), and listed in Osterbrock \& Martel (1993) or 
Ho et al.~(1995).   When classifications differed we listed the classification
on the left by Maiolino \& Rieke followed by that of Ho et al.~on
the right;  
(4) Heliocentric velocity in km/s;
(5) HST Proposal ID for the 1.6 micron images followed
by the NICMOS camera number used to image the galaxy;
(6) Best fitting surface brightness profile shape in the central arcsecond.  
We fit the sum of an unresolved nuclear source and either an
exponential profile ($S(r) = S_0 e^{-{r/h}}$) or a powerlaw profile
($S(r) = S_{r=1''} r^{-\alpha}$).
P refers to the powerlaw and E refers to the exponential.  When an unresolved
nuclear component dominates the central arcsecond 
we do not record the profile shape.
(6) The flux of the unresolved component in mJy.
Errors are estimated based on the difference between nuclear fluxes estimated 
from the best unresolved+exponential profile fit and
that of the best unresolved+powerlaw profile fit.  
We restricted the estimated error to be larger than 10\% of the unresolved
component estimated from the best fit.
When the galaxy was resolved we list as an upper limit the flux of
an unresolved component which was the result of 
the best fitting unresolved+exponential profile fit.
(7)  When the galaxian profile fit was an exponential we record
the central surface brightness $S_0$, in mJy/arcsecond$^2$.
Whent the galaxian profile shape was a power law we record the surface
brightness, $S_{r=1''}$, at a radius of $1''$.
(8) When the galaxian profile fit was an exponential we record
exponential scale length, $h$, of the galaxian component in arcseconds.
When the galaxian profile fit was a powerlaw we record
the exponent $\alpha$.
When the point source was extremely bright, the galaxian
profile was not constrained so $S$ and $h$ or $\alpha$ are not listed.
%
}
\end{deluxetable}

\begin{deluxetable}{lccccccccccc}
\tablewidth{0pt}
\tablecaption{Additional Seyferts which are in the CfA sample}    
\tablehead{
\multicolumn{1}{l}{Galaxy}     &
\colhead{Nucleus}    &
\colhead{Type}       &
\colhead{Proposid}   &
\colhead{$v_{hel}$}  &
\colhead{profile}    &
\colhead{$f_{nuc}$}  &
\colhead{$S_0|S_{r=1''}$}        &
\colhead{$h|\alpha$}  \\
\colhead{}                &
\colhead{}                &
\colhead{}                &
\colhead{}                &
\colhead{km/s}            &
\colhead{}                &
\colhead{mJy}             &
\colhead{mJy $''^{-2}$}     &
\colhead{}          \\
\colhead{(1)}             & 
\colhead{(2)}             & 
\colhead{(3)}             & 
\colhead{(4)}             & 
\colhead{(5)}             & 
\colhead{(6)}             & 
\colhead{(7)}             & 
\colhead{(8)}             & 
\colhead{(9)}              
} 
\startdata
%
NGC 1144&: & 2    & 7867/1 & 8648  & E &   0.08 $\pm$   0.08 &   8.56 &   0.38 \\
NGC 5252&F & 1.9  & 7330/2 & 6926  & P &   0.60 $\pm$   0.21 &   1.34 &   0.80 \\
NGC 5283&- & 2    & 7867/1 & 2700  & P &   $<0.3$            &   1.99 &   0.91 \\
NGC 5695&- & 2    & 7867/1 & 4225  & P &   $<0.2$            &   1.65 &   0.72 \\
NGC 5929&- & 2    & 7330/2 & 2492  & P &   $<0.25$           &   1.64 &   0.81 \\
NGC 5940&* & 1    & 7328/1 & 10115 & E &   0.93 $\pm$   0.21 &   4.21 &   0.34 \\
NGC 6104&* & 1.5  & 7328/1 & 8382  & P &   0.13 $\pm$   0.05 &   0.45 &   0.75 \\
NGC 7674&* & 2    & 7328/1 & 8713  & E &   4.39 $\pm$   0.44 &  13.57 &   0.22 \\ 
NGC 7682&- & 2    & 7867/1 & 5134  & P &   $<0.01$           &   0.75 &   0.77 \\
MRK 231 &* & 1    & 7213/2 & 12642 & - &  80.1  $\pm$   8.0  &   -    &   -    \\
MRK 266 &- & 2    & 7328/2 & 8360  & P &    $<0.1$           &   1.85 &   0.48 \\
MRK 334 &* & 1.8  & 7867/1 & 6582  & E &   6.83 $\pm$   0.86 &  60.62 &   0.11 \\
MRK 461 &- & 2    & 7867/1 & 4856  & P &    $<0.4$           &   0.63 &   1.21 \\
MRK 471 &* & 1.8  & 7328/1 & 10263 & E &   0.32 $\pm$   0.04 &   3.54 &   0.52 \\
MRK 573 &F & 2    & 7330/2 & 5174  & P &   0.54 $\pm$   0.23 &   1.90 &   0.73 \\
UGC 6100&- & 2    & 7867/1 & 8778  & P &    $<0.15$          &   0.87 &   0.92 \\
UGC 12138&*& 1.8  & 7328/1 & 7375  & E &   2.59 $\pm$   0.26 &  10.50 &   0.37 \\
UM 146  &* & 1.9  & 7328/1 & 5208  & E &   0.82 $\pm$   0.19 &   5.64 &   0.29 \\
%
\enddata
\tablecomments{
See Table 1 for more information.
}
\end{deluxetable}

\begin{deluxetable}{lccccccccccc}
\tablewidth{0pt}
\tablecaption{Additional Seyferts}        
\tablehead{
\multicolumn{1}{l}{Galaxy}     &
\colhead{Nucleus}    &
\colhead{Type}       &
\colhead{Proposid}   &
\colhead{$v_{hel}$}  &
\colhead{profile}    &
\colhead{$f_{nuc}$}  &
\colhead{$S_0|S_{r=1''}$}        &
\colhead{$h|\alpha$}  \\
\colhead{}                &
\colhead{}                &
\colhead{}                &
\colhead{}                &
\colhead{km/s}            &
\colhead{}                &
\colhead{mJy}             &
\colhead{mJy $''^{-2}$}     &
\colhead{}          \\
\colhead{(1)}             & 
\colhead{(2)}             & 
\colhead{(3)}             & 
\colhead{(4)}             & 
\colhead{(5)}             & 
\colhead{(6)}             & 
\colhead{(7)}             & 
\colhead{(8)}             & 
\colhead{(9)}              
} 
\startdata
%
NGC 1672    &- & 2    & 7330/2 & 1350  & P &   $<1.1$            &   4.80 &   0.83 \\
NGC 3079$^h$&- & 2    & 7330/2 & 1125  & P &   $<0.1$            &   7.43 &   0.24 \\
NGC 3486$^h$&F & 2    & 7330/2 & 681   & P &   0.78 $\pm$   0.31 &   1.68 &   0.83 \\
NGC 3718$^h$&: & L1.9 & 7330/2 & 994   & P &   0.98 $\pm$   0.61 &   4.89 &   0.73 \\
NGC 4117    &F & 2    & 7330/2 & 958   & P &   0.24 $\pm$   0.18 &   1.07 &   0.81 \\
NGC 4303$^h$&* & 2/H  & 7330/2 & 1566  & P &   3.72 $\pm$   0.37 &   3.84 &   0.58 \\
NGC 4725$^h$&- & 2    & 7330/2 & 1206  & P &   $<0.7$            &   4.60 &   0.79 \\
NGC 4968    &- & 2    & 7330/2 & 2957  & P &   $<0.4$            &   1.33 &   1.08 \\
NGC 6951$^h$&- & 2    & 7330/2 & 1424  & P &   $<0.3$            &   2.55 &   0.80 \\
MRK 1210    &* & 2    & 7330/2 & 4046  & E &   1.51 $\pm$   0.15 &   6.47 &   0.41 \\
MRK 78      &- & 2    & 7330/2 &11137  & P &   $<0.6$            &   1.44 &   1.07 \\
MRK 477     &F & 2    & 7330/2 &11332  & P &   0.29 $\pm$   0.16 &   0.31 &   1.10 \\
ESO138-G1   &* & 2    & 7330/2 & 2740  & P &   5.36 $\pm$   0.67 &   1.89 &   0.23 \\
ESO137-G34  &- & 2    & 7330/2 & 2747  & P &   $<0.1$            &   1.44 &   1.07 \\  
ESO362-G08&-& 2   & 7328/1 & 4785  & P &   $<0.3$            &   4.32 &   0.70 \\
IRAS1443+2714&*& 2& 7328/1 & 8814  & E &   1.96 $\pm$   0.20 &   5.57 &   0.28 \\
IRAS2302-0004&*& 2& 7328/2 & 7585  & E &   3.82 $\pm$   0.38 &   7.89 &   0.23 \\
IRAS1833-6528&-& 2& 7328/2 & 3983  & P &   $<0.3$            &   1.95 &   0.87 \\  
NGC 1019&* &  1   & 7328/1 & 7251  & E &   1.27 $\pm$   0.16 &   7.29 &   0.27 \\
NGC 7319&: &  2   & 7328/2 & 6764  & P &   0.07 $\pm$   0.02 &   0.68 &   0.73 \\
MRK 1   &- &  2   & 7328/1 & 4780  & P &   $<0.2$            &   0.40 &   1.23 \\
MRK 6   &* &  1.5 & 7328/1 & 5537  & E &  30.6  $\pm$   3.1  &   9.29 &   0.23 \\
MRK 40  &* &  1   & 7328/1 & 6323  & E &   0.78 $\pm$   0.18 &   6.59 &   0.27 \\
MRK 42  &* &  1   & 7328/1 & 7200  & E &   1.35 $\pm$   0.25 &   4.56 &   0.23 \\
MRK 176 &* &  2   & 7328/2 & 8346  & E &   2.96 $\pm$   0.30 &  16.34 &   0.31 \\
MRK 372 &* &  1.5 & 7328/1 & 9300  & E &   0.69 $\pm$   0.10 &  14.51 &   0.27 \\
MRK 493 &* &  1   & 7328/1 & 9569  & P &   4.74 $\pm$   0.47 &   0.23 &   1.23 \\
MRK 516 &: &  1.8 & 7328/1 & 8519  & E &   0.18 $\pm$   0.10 &  15.82 &   0.22 \\
MRK 915 &* &  1   & 7328/1 & 7228  & P &   4.25 $\pm$   0.43 &   0.59 &   0.97 \\
MRK 1048&* &  1   & 7328/2 & 12934 & P &  12.2  $\pm$   1.2  &   0.39 &   1.05 \\
MRK 1261&F &  1   & 7328/1 & 7808  & P &   0.18 $\pm$   0.02 &   1.31 &   0.69 \\
MRK 1308&* &  2   & 7328/2 & 1087  & E &   0.89 $\pm$   0.21 &   1.18 &   0.62 \\
IC 4870 &* &  2   & 7328/2 & 889   & E &   0.62 $\pm$   0.17 &   2.14 &   0.15 \\
UM 625  &* &  2   & 7328/1 & 7495  & E &   0.23 $\pm$   0.03 &   2.42 &   0.60 \\
NGC 4472$^h$&- &  2   & 7453/2 &  868  & P &   $<0.05$           & 0.04   &$\sim 0.0$\\
NGC 4565$^h$&F &  1.9 & 7331/2 & 1282  & P &   0.63 $\pm$   0.24 &   6.55 &   0.44 \\
IC 4329A   & * &  1.2 & 7172/2 & 4813  & - &  59.4  $\pm$   5.9  &   -    &   -    \\
%
%
\enddata
\tablecomments{
This table lists additional Seyferts from proposals 7330, 7328,
from the tabulation of Ho et al.~(1995) and the Seyfert 1 galaxy
IC~4329A.
See Table 1 for more information.
}
\end{deluxetable}

\begin{deluxetable}{lccccccccccc}
\tablewidth{0pt}
\tablecaption{Control galaxies with unresolved nuclear sources}         
\tablehead{
\multicolumn{1}{l}{Galaxy}     &
\colhead{Nucleus}    &
\colhead{Type}       &
\colhead{Proposid}   &
\colhead{$v_{hel}$}  &
\colhead{profile}    &
\colhead{$f_{nuc}$}  &
\colhead{$S_0|S_{r=1''}$}        &
\colhead{$h|\alpha$}  \\
\colhead{}                &
\colhead{}                &
\colhead{}                &
\colhead{}                &
\colhead{km/s}            &
\colhead{}                &
\colhead{mJy}             &
\colhead{mJy $''^{-2}$}     &
\colhead{}          \\
\colhead{(1)}             & 
\colhead{(2)}             & 
\colhead{(3)}             & 
\colhead{(4)}             & 
\colhead{(5)}             & 
\colhead{(6)}             & 
\colhead{(7)}             & 
\colhead{(8)}             & 
\colhead{(9)}              
} 
\startdata
%
%
NGC 157 &F &      & 7330   & 1668  & P &   0.29 $\pm$   0.08 &   1.75 &   0.24 \\
NGC 404 &* & L2   & 7330   & -48   & P &   2.99 $\pm$   1.81 &   1.76 &   1.19 \\
NGC 578 &* &      & 7330   & 1630  & E &   0.24 $\pm$   0.02 &   0.71 &   1.42 \\ 
NGC 864 &: & H    & 7330   & 1562  & P &   0.31 $\pm$   0.06 &   0.72 &   0.59 \\
NGC 1530&F &      & 7330   & 2461  & P &   0.25 $\pm$   0.02 &   1.67 &   0.24 \\
NGC 2776&F & H    & 7330   & 2626  & P &   0.16 $\pm$   0.03 &   0.68 &   0.53 \\
NGC 4030&F &      & 7330   & 1460  & P &   0.43 $\pm$   0.11 &   2.00 &   0.53 \\
NGC 4380&F & H    & 7330   & 967   & P &   0.13 $\pm$   0.08 &   0.56 &   0.78 \\
NGC 5383&: & H    & 7330   & 2250  & P &   0.07 $\pm$   0.04 &   1.24 &   0.43 \\
NGC 5970&F & L2   & 7330   & 1957  & E &   0.09 $\pm$   0.01 &   1.73 &   1.58 \\
NGC 6384&: & T2   & 7330   & 1665  & P &   0.17 $\pm$   0.05 &   1.96 &   0.38 \\
NGC 6412&: & H    & 7330   & 1324  & P &   0.09 $\pm$   0.01 &   0.27 &   0.26 \\
NGC 7126&: &      & 7330   & 2981  & E &   0.22 $\pm$   0.10 &   5.60 &   0.47 \\
\enddata
\tablecomments{
See Table 1 for more information.  All these images were observed
with Camera 2.  When possible, classifications from Ho et al.~(1995) 
are shown.
}
\end{deluxetable}

\begin{deluxetable}{lccccccccccc}
\tablecaption{Control galaxies lacking unresolved emission}
\tablewidth{0pt}
\tablehead{
\multicolumn{1}{l}{Galaxy}     &
\colhead{$v_{hel}$}  &
\colhead{Profile}    &
\colhead{$S_0|S_{r=1''}$} &
\colhead{$h|\alpha$} &
}
\startdata
%
NGC 214  &   4534  & P & 1.09 &   0.75 \\ 
NGC 357  &   2406  & P & 2.70 &   0.66 \\
NGC 628  &    657  & P & 0.20 &   1.28 \\
NGC 1300 &   1568  & P & 1.52 &   0.96 \\
NGC 1398 &   1407  & P & 6.63 &   0.64 \\
NGC 1638 &   3320  & P & 2.82 &   0.60 \\
NGC 1961 &   3934  & P & 2.55 &   0.54 \\
NGC 2179 &   2798  & E & 9.01 &   0.68 \\
NGC 2223 &   2722  & P & 9.01 &   0.68 \\
NGC 2336 &   2204  & P & 1.55 &   0.69 \\
NGC 2460 &   1442  & P & 1.91 &   0.84 \\
NGC 2841 &   638   & P & 8.95 &   0.63 \\
NGC 2903 &   556   & P & 1.63 &   0.92 \\
NGC 2985 &   1322  & P & 3.39 &   0.84 \\
NGC 3032 &   1533  & P & 1.19 &   1.52 \\
NGC 3145 &   3506  & P & 2.73 &   0.52 \\
NGC 3300 &   3045  & P & 1.27 &   0.78 \\
NGC 3351 &   778   & P & 2.65 &   0.39 \\
NGC 3368 &   897   & P & 5.95 &   0.83 \\
NGC 3458 &   1818  & P & 3.67 &   0.74 \\
NGC 3627 &   897   & P & 5.88 &   0.78 \\
NGC 3630 &   727   & P & 4.92 &   0.75 \\
NGC 3865 &   1509  & E & 10.07&   0.32 \\
NGC 4143 &   985   & P & 6.74 &   0.70 \\
NGC 4254 &   2407  & P & 0.69 &   1.15 \\
NGC 4260 &   1958  & P & 1.44 &   0.67 \\
NGC 4314 &   963   & P & 2.75 &   0.67 \\
NGC 5054 &   1741  & P & 2.11 &   0.88 \\
NGC 5064 &   3002  & P & 4.45 &   0.55 \\
NGC 5614 &   3892  & P & 3.64 &   0.79 \\
NGC 5691 &   1870  & P & 0.21 &   0.54 \\ 
NGC 5739 &   5377  & P & 4.31 &   0.69 \\
NGC 6744 &   841   & P & 2.69 &   0.77 \\
NGC 6946 &   48    & E & 21.55&   0.86 \\
NGC 7096 &   3100  & E & 12.06&   0.56 \\
NGC 7177 &   1150  & E & 8.81 &   0.84 \\
NGC 7392 &   3192  & P & 1.90 &   0.86 \\ 
NGC 7716 &   2571  & P & 2.29 &   0.74 \\     
NGC 7744 &   3098  & P & 4.66 &   0.80 \\
NGC 7814 &   1050  & E &10.91 &   0.94 \\
IC 5267  &   1713  & P & 3.70 &   0.76 \\
%
\enddata
\tablecomments{
All galaxies were part of proposal 7330 and were observed
with NICMOS Camera 2.
See Table 1 for more information.
}
\end{deluxetable} 

\clearpage

\begin{table}
\caption{Numbers of Seyfert Galaxies with point sources \label{key}}
\vskip 0.1truein
\begin{tabular}{|l|cccc|c|cccc|cc|c|cccc|c|}
\tableline
\tableline
Sample   &\multicolumn{5}{c|}{2.0} &\multicolumn{4}{c|}{1.8-1.9}& 
\multicolumn{2}{c|}{1.0-1.5}& Total Sy& \multicolumn{5}{c|}{Non-Sy}    \\
\tableline
       & *  & F & :& - &\%Res  & * & F& :& -&*  & F &      & * &F &: & -&\%Res\\
\tableline
CfA    & 4  &3  &1 &7  &$53|47$&9  &1 &0 &0 &10 &0  &  35  &   &  &  &  &  \\
RSA    & 14 &5  &6 &14 &$64|49$&8  &0 &0 &1 &9  &0  &  57  &   &  &  &  &  \\
7330   & 13 &8  &3 &19 &$56|49$&5  &1 &1 &0 &4  &0  &  54  & 2 &6 &5 &41&$24|15$\\
Ho     & 0  &1  &2 & 6 &$33|11$&3  &2 &1 &2 &9  &0  &  26  &   &  &  &  &  \\
All Sy & 24 &9  &8 &32 &$56|45$&12 &2 &2 &1 &21 &1  & 112  &   &  &  &  &  \\
\tableline
\end{tabular}
\tablecomments{The types of nuclei seen (*, F, :, -) are described in 
the caption to Table 1.   We list here the number of each
type of Seyfert galaxy with each type of nucleus.  For the Seyfert 2 
and non-Seyfert samples we also list in the column labelled \%Res
the percent of Seyferts of that type with emission from
an unresolved component (type *, F or :) and that from
an unresolved component at high confidence (type * and F).
The subsets considered here are the Seyfert 
galaxies which are part of the CfA and RSA samples, 
galaxies from HST proposal 7330 including Seyfert and non-Seyfert samples,
Seyferts which were also tabulated by Ho et al.~(1995), 
and all the Seyferts we have listed in Tables 1-3.
We have used classifications listed in Maiolino \& Rieke (1995)
except for the Ho et al.~sample in which case we use their classifications.
All Seyfert 1.0-1.5 galaxies listed Tables 1-3 display prominent
unresolved nuclear sources.  Almost all Seyfert 1.8 or 1.9 galaxies
display some level of unresolved emission.
About 50\% of the Seyfert 2 galaxies display significant unresolved
emission compared to $\sim 25\%$ of the normal galaxies from proposal 7330.
Proposal 7330 contains a non-Seyfert sample with distribution that is intended
to match that of its Seyfert sample.
}
\end{table}

\begin{deluxetable}{lcccccc}
\tablewidth{0pt}
\tablecaption{Correlation Statistics}
\tablehead{
\colhead{Type}        &
\colhead{Quantities}  &
\colhead{$N$}         &
\colhead{$R_s$}       &
\colhead{Probability} 
}
\startdata
Sy1.0-1.9 & [OIII] vs. $1.6\micron$& 26 &   0.67 & $>0.99$ \\
Sy2.0     & [OIII] vs. $1.6\micron$& 31 &   0.22 & $0.75$  \\
Sy1.0-1.9 & HX vs.     $1.6\micron$& 14 &   0.78 & $>0.99$ \\
Sy2.0     & HX vs.     $1.6\micron$& 18 &   0.34 & $0.83$  \\
\enddata
\tablecomments{Spearman rank-order correlation coefficient, $R_s$
for Seyferts shown in Figure 3.  The probability of a correlation
between fluxes (not luminosities) are shown in the right most column.  
The number of objects is given by $N$.
}
\end{deluxetable}